\documentclass[aps,pra,a4paper,twocolumn,amsmath,amssymb,showpacs,floatfix]{revtex4}

\usepackage{graphicx}
\usepackage{epstopdf}

\newcommand{\Op}[1]{\boldsymbol{\mathsf{\hat{#1}}}}

\newcommand{\half}{\frac{1}{2}}

\begin{document}
\hyphenation{Fesh-bach}

\title{Perspectives for coherent optical formation of strontium molecules in
  their electronic ground state}

\date{\today}
\author{Christiane P. Koch}
\email{ckoch@physik.fu-berlin.de}
\affiliation{Institut f\"ur Theoretische Physik,
  Freie Universit\"at Berlin, Arnimallee 14, D-14195
  Berlin, Germany}


\begin{abstract}
  Optical Feshbach resonances [Phys. Rev. Lett. 94, 193001 (2005)]
  and pump-dump photoassociation with
  short laser pulses [Phys. Rev. A 73, 033408 (2006)]
  have been proposed as means to coherently form stable
  ultracold alkali dimer molecules.
  In an optical Feshbach resonance, the intensity and possibly
  frequency of a cw laser are ramped up linearly followed by a sudden
  switch-off of the laser. This is applicable to tightly trapped atom
  pairs. In short-pulse photoassociation, the pump pulse forms a
  wave-packet in an  electronically excited state. The ensuing dynamics
  carry the wave-packet to shorter internuclear distances where, after
  half a vibrational period, it can be deexcited to the electronic
  ground state by the dump pulse. Short-pulse photoassociation is 
  suited for both shallow and tight traps. 
  The applicability of these two means to produce ultracold molecules
  is investigated here for $^{88}$Sr. Dipole-allowed transitions proceeding
  via the $B^1\Sigma_u^+$ excited state as well as transitions near
  the intercombination line are studied.
\end{abstract}

\maketitle

%--------------------------------------------------------------------------------%
\section{Introduction}
\label{sec:intro}

The cooling and trapping of alkaline-earth metals and systems with
similar electronic structure such as ytterbium
have been the subject of intense research over the last decade. 
The interest in ultracold group II atoms
was triggered by the quest for new optical
frequency standards \cite{LudlowSci08}. 
The extremely narrow linewidth of the intercombination transition,
together with the magic wavelength of an optical
lattice \cite{KatoriNat05}, is at the heart of the clock proposals. 
This narrow width has several interesting implications, such as
a very low Doppler temperature for laser cooling \cite{BinnewiesPRL01},
or optical Feshbach resonances 
involving small losses and thus holding the
promise of easy optical control
\cite{CiuryloPRA05,CiuryloPRA06,PascalPRA06}. 
Additionally ultracold alkaline-earth metals offer many more exciting
perspectives: They are candidates for
high precision measurements \cite{FerrariPRL06,ZelevinskyPRL08},
for studies of ultracold mixtures \cite{Goerlitz07},  for
the realization of quantum information processing \cite{HayesPRL07,Daley08} 
and quantum phases with unusual symmetries \cite{HemmerichPRL07}. 

\begin{figure}[b]
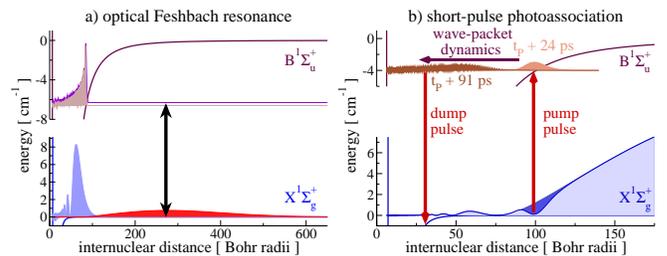

  \centering
  \includegraphics[width=0.49\linewidth]{fig1a} %{optFR}
  \includegraphics[width=0.49\linewidth]{fig1b} %pulsePA}
  \caption{(Color online) Coherent formation of ultracold strontium molecules in
    their electronic ground state via (a) an optical Feshbach
    resonance where an atom pair in the lowest state of a tight trap
    is adiabatically compressed to a molecule 
    and (b) by short pulse photoassociation where molecules are formed
    in a pump-dump sequence. 
    }
  \label{fig:scheme}
\end{figure}

The present study focusses on strontium and addresses the question of
how to produce dimer molecules in their electronic ground state
from ultracold $^{88}$Sr atoms. Emphasis is put on a coherent
formation mechanism. Two different time-dependent schemes are
studied, a ramp over an optical Feshbach
resonance \cite{MyPRL05} and a 
sequence of short pump-dump laser pulses \cite{MyPRA06a},
cf. Fig.~\ref{fig:scheme}, 
that are both based on photoassociation resonances.

Photoassociation, the excitation of two colliding atoms into a weakly
bound molecular level of an electronically excited state by laser light,
has emerged as a standard tool  to study the scattering properties of
ultracold atoms \cite{JonesRMP06}. For alkaline-earth
species, both dipole allowed transitions and transitions near the
intercombination line can be employed. Photoassociation based on
dipole-allowed transitions near the $^1S_0- \,^1P_1$ atomic resonance
was observed  for calcium \cite{ZinnerPRL00}, for strontium 
\cite{NagelPRL05,MickelsonPRL05}, and for ytterbium
\cite{TakasuPRL04}.
Near the narrow-line atomic intercombination transition, photoassociation
spectroscopy has been performed for strontium
\cite{ZelevinskyPRL06,Martinez08} 
and for ytterbium \cite{TojoPRL06}.

In addition to gathering spectroscopic information, photoassociation
serves as a method to produce stable ultracold molecules \cite{FiorettiPRL98}.
Depending on the properties of the electronically excited state,
spontaneous decay may yield molecules in their 
electronic ground state \cite{DionPRL01}.  
However, spontaneous decay toward molecules
instead of a pair of atoms occurs only 
if the vibrational wavefunctions show a large probability amplitude at
short internuclear distances $R$. On the other hand, 
large free-bound Franck-Condon factors and hence very extended
vibrational wavefunctions are required for photoassociation.
Ideally the vibrational wavefunctions  should therefore consist
of both a short-range part and a long-range part.
In heavy alkali dimers such wavefunctions exist thanks to strong
spin-orbit coupling \cite{DionPRL01}. In particular, the softly
repulsive part of a purely long-range well \cite{FiorettiPRL98}
and resonant spin-orbit coupling
\cite{DionPRL01,HyewonMyPRA07,FiorettiJPB07} can provide the
necessary '$R$-transfer' mechanism. If the vibrational wavefunctions of
the electronically excited state do not possess this dual character,
molecules in the last one or two levels of the electronic
ground state can be formed at most. Overwhelmingly, spontaneous decay will
simply redissociate the molecules to a pair of atoms. 

Photoassociation as a means to produce molecules in their electronic
ground state may be more efficient if time-dependent laser fields are
employed. For example a sequence of two laser pulses, a pump pulse and
a time-delayed dump pulse can be utilized \cite{MyPRA06a} (see also
Fig.~\ref{fig:scheme}b). The pump pulse
excites a vibrational wave-packet in the electronically excited state.
The wave-packet moves to shorter internuclear distances where it can be transferred
to the electronic ground state by the dump pulse. 
The character of the vibrational wavefunctions  decides the fate of
molecule formation: The wave-packet dynamics 
lead to appreciable overlap with molecular levels in the electronic
ground state only if the eigenfunctions from which the wave-packet is
built provide for it \cite{MyPRA06b}. This reflects the weak field approach
implicit in the pump-dump scheme where the maximum transition probabilities
are completely determined by the Franck-Condon factors and the spectral
amplitudes of the field. 

An alternative time-dependent scheme for creating molecules in their
electronic ground state consists in ramping the intensity and possibly
also frequency of a continuous-wave (cw) laser over a photoassociation
resonance \cite{MyPRL05} (see also Fig.~\ref{fig:scheme}a). This has
been termed optical Feshbach 
resonance (FR), in analogy to the formation of molecules with magnetic
fields \cite{KoehlerRMP06}. It requires an initial condition where
two atoms are not too far from each other such as in a Mott insulator
state of an optical lattice with two atoms per lattice site. The
intensity of the laser field is ramped up slowly while the
frequency of the field is chosen such that the population of the
excited state is kept minimal \cite{MyPRL05}. This ensures an adiabatic
compression of the wavefunction while avoiding loss due to spontaneous
emission. In 
the case of rubidium, a molecule formation efficiency of 50\% was
predicted for realistic laser parameters.

While  the pump-dump sequence with short pulses and the ramp over an
optical FR with a cw laser both represent  \textit{coherent}
molecule formation schemes, they differ in a number of aspects. (1)
The timescale for the pump-dump sequence is determined by the
vibrational periods in the electronically excited state which are on
the order of 100$\,$ps. Ramping over an optical FR
needs to be done much more slowly,
typically over $0.1 - 1\,\mu$s. This timescale is set by the initial
state, i.e. by the frequency of the trapping potential. 
(2) Short pulses address only a small part of the initial wavefunction
in a certain range of internuclear distances, the photoassociation
window. When ramping over an 
optical FR, the whole wavefunction is adiabatically
deformed. 

Here, the two schemes are applied to the formation of strontium
dimers in their electronic ground state. They are schematically
depicted for optically allowed transitions proceeding via the Sr$_2$
$B^1\Sigma_u^+$ excited state in Fig.~\ref{fig:scheme}.
Whether coherent molecule formation by optical transitions near the
intercombination is possible, is also studied.
Similarities and differences between alkali and alkaline earth dimers
are analyzed by comparing $^{88}$Sr$_2$ to $^{87}$Rb$_2$. 
The paper is organized as
follows. Section~\ref{sec:model} outlines the two-state model that is required
to describe molecule formation by optically allowed transitions as
well as the three-state model that presents a simplified, yet sufficiently
accurate view of the physics near the intercombination line.
The optical FRs are studied in Section~\ref{sec:OFR}, short-pulse
PA is discussed in Section~\ref{sec:PA}, and
Section~\ref{sec:concl} conludes.

\section{Model}
\label{sec:model}

Two strontium atoms initially interacting via the electronic
ground state potential are considered. Assuming a harmonic trapping
potential, the center of mass motion can be 
integrated analytically and is omitted from the description.
 For sufficiently low temperatures only $s$-wave encounters need to be
 taken into account. The electronic ground state potential
 supports both bound levels and scattering states. A laser can excite
an unbound atom pair via an  electric dipole-allowed transition
 into bound molecular levels below the $^1S_0+^1P_1$ asymptote, 
 or via transitions near the intercombination line into bound
 molecular levels below the $^1S_0+^3P_1$ asymptote. 

The Hamiltonian modelling electric dipole-allowed transitions between
the $X^1\Sigma_g^+$ 
electronic ground state and the $B^1\Sigma_u^+$ electronically excited
state is given by 
\begin{equation}
  \label{eq:H_dip}
  \Op{H}^{DA}(t) = 
  \begin{pmatrix}
    \Op{H}_g && \Op{\mu} \,E(t) \\
     \mu \,E(t)  && 
    \Op{H}_e - \Delta_L(t) - \frac{i\hbar}{2}\Gamma
  \end{pmatrix} \,,
\end{equation}
where the rotating-wave approximation (RWA) is invoked.
The single-channel Hamiltonians, $\Op{H}_{g(e)}$, consist of the kinetic
energy, the molecular potential energy curves and the trap
potential, 
$\Op{H}_{g(e)} = \Op{T} + V_{g(e)}(\Op{R}) +(-)
V_\mathrm{tr}(\Op{R})$.
The potentials $V_{g(e)}$ are schematically shown in
Fig.~\ref{fig:scheme}, they are taken from Refs.~\cite{Martinez08} ($V_g$)
and \cite{NagelPRL05} ($V_e$).
The trap is assumed to be harmonic,
$V_\mathrm{tr}(\Op{R}) = \half m \omega_\mathrm{tr}^2 \Op{R}^2$ with
$m$ denoting the reduced mass and $\omega_\mathrm{tr}$ the trapping frequency.

Following Ref.~\cite{ZelevinskyPRL06},
a three-channel Hamiltonian, again in the RWA,
is used to model transitions near the intercombination line,
\begin{widetext}
\begin{equation}
  \label{eq:H_IC}
  \Op{H}^{IC}(t) = 
  \begin{pmatrix}
    \Op{H}_g && \Op{\tilde\mu} \,E(t) && 2 \Op{\tilde\mu} \,E(t) \\
     \Op{\tilde\mu} \,E(t)  &&  \Op{H}_{0_u^+} + 4 V_{rot}(\Op{R}) -
     \Delta_L(t) - \frac{i\hbar}{2}\tilde\Gamma &&
     -\frac{1}{2\sqrt{2}} V_{rot}(\Op{R})
     \\
     2\Op{\tilde\mu} \,E(t)  && -\frac{1}{2\sqrt{2}} V_{rot}(\Op{R})
     && \Op{H}_{1_u} + 2V_{rot}(\Op{R}) - \Delta_L(t) - \frac{i\hbar}{2}\tilde\Gamma
  \end{pmatrix} \,,
\end{equation}
\end{widetext}
with the rotational couplings given in terms of
$V_{rot}(\Op{R})=\frac{\hbar^2}{2 \mu \Op{R}^2}$.
The potentials of Lennard-Jones type fitted to experimental data 
are found in Ref.~\cite{ZelevinskyPRL06}.

The transition dipole operators are assumed to be independent of $R$ and
fixed at their atomic values, i.e. $\mu$ and $\tilde\mu$ are given in
terms of the respective 
$C_3$ coefficients, cf. Table~\ref{tab:const}.
\begin{table}[tb]
  \centering
  \begin{tabular}{|c|c|c|}
    \hline
    & near the intercombination line & dipole allowed \\ \hline
    $C_3$ & 0.0075$\,E_h\,a_0^3$ & 18.54$\,E_h\,a_0^3$ \\
    lifetime & 21.46 $\mu$s & 5.22 ns \\ \hline
  \end{tabular}
  \caption{$C_3$ coefficient and lifetime of strontium atoms for
    electronic states accessed by transitions near the
    intercombination line and by dipole-allowed transitions}
  \label{tab:const}
\end{table}
The electric field amplitude is denoted by $E(t)$.
In case of the optical FR it corresponds to a linear function of time
or to a constant, while a Gaussian pulse envelope is assumed for the
short-pulse photoassociation.  
The detuning, $\Delta_L(t) = \hbar (\omega_\mathrm{at}-\omega_L(t))$,
becomes time-dependent only in the case of the optical FR, and for 
simplicity, a linear time-dependence of $\Delta_L(t)$ is chosen. The
finite lifetime of the excited states is modelled by an exponential
decay with decay rates $\Gamma$, $\tilde\Gamma$. This corresponds to
the assumption that every excited state molecule which undergoes
spontaneous emission yields a hot atom pair which is lost from the trap. 
The lifetimes corresponding to $\Gamma$, $\tilde\Gamma$ 
are listed in Table~\ref{tab:const}.

The Hamiltonians are each represented on a Fourier grid employing an
adaptive grid step
\cite{SlavaJCP99,WillnerJCP04,ShimshonCPL06}. Setting $E(t)=0$, the
wavefunctions and binding energies of the vibrational levels are
obtained by diagonalization.
Analogously, dressed states are calculated by diagonalization
for fixed values of $E(t)$ and $\Delta_L(t)$.

Coherent optical molecule formation is
investigated by solving the time-dependent Schr\"odinger equation,
\[
\frac{\partial}{\partial t} |\Psi(t)\rangle = \Op{H}(t)  |\Psi(t)\rangle,
\]
with a Chebychev propagator \cite{RonnieReview94} where $\Op{H}(t)$
corresponds to either Eq.~(\ref{eq:H_dip}) or to Eq.~(\ref{eq:H_IC}). 
The state of the system at time $t$, $|\Psi(t)\rangle$, is represented
on the mapped Fourier grid 
for each channel $i$, $\Psi_i(R;t)=\langle R |
\Psi_i(t)\rangle$. 

\section{Strontium molecule formation via Optical Feshbach resonance}
\label{sec:OFR}

Molecule formation via an optical FR is based on the formal analogy between a
photoassociation resonance and a magnetic FR. Instead of ramping the
magnetic field magnitude, the intensity and detuning of a cw-laser are
increased in order to cross the resonance \cite{MyPRL05}. Then,
in the dressed-state picture, the number of eigenstates below the
ground state dissociation limit is increased by one. If the ramp is performed
slowly enough, the lowest trap state is adiabatically connected to the
dressed state just below the ground state dissociation limit. This
dressed state is an eigenstate of the Hamiltonian for the final values
of field amplitude and detuning. However, the excited state components of
the state of the system are subject to decay due to spontaneous
emission. Therefore the laser coupling between the ground and excited
state should be switched off before significant spontaneous emission
losses occur. A sudden switch-off of the coupling corresponds to a
projection of the dressed state onto the eigenstates of the
field-free Hamiltonian, i.e. onto the vibrational levels. If the
dressed state before the switch-off has a large overlap with the last
bound level of the electronic ground state potential, molecules are
formed.

The success of this scheme is determined by the
value of the dressed-state projections onto the last bound level which
depends on the laser intensity and detuning, and
by the time within which an adiabatic ramp can be performed. This time
is constrained from above and below by the requirement of avoiding
spontaneous emission loss and by the requirement of adiabaticity,
respectively.
Both limiting timescales are controlled by experimentally accessible
parameters. The loss due to spontaneous emission depends on how much
population is actually transferred to the electronically excited state --
this is determined by the laser intensity and detuning.
On the other hand, adiabaticity
requires the ramp to be slower than the slowest internuclear dynamics,
that is slower than the vibrational period of the lowest trap state --
this is determined by the trap frequency, i.e. the intensity of the
trapping laser. Vibrational periods in the sub-$\mu$s regime are
realized in tight traps such as a deep optical lattice.

\subsection{Dipole allowed transitions $X^1\Sigma_g^+ \longrightarrow B^1\Sigma_u^+$ }

\begin{figure}[tb]
  \centering
  \includegraphics[width=0.9\linewidth]{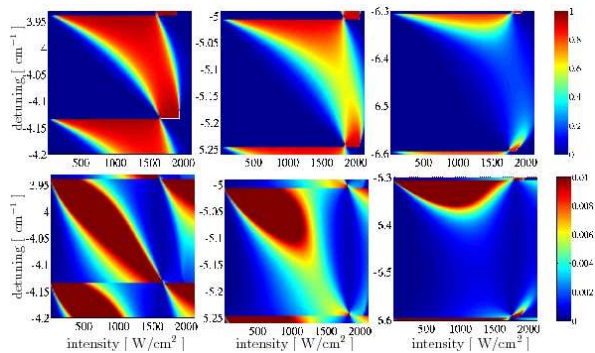} %{plast_pope}  
  \caption{(Color online)
    Excited state population (bottom) and
    projection onto the last bound
    level (top) as a function of laser intensity and detuning
    for three different pairs of photoassociation resonances
  }
  \label{fig:FR}
\end{figure}

The two quantities characterizing gain and loss in molecule formation
via an optical FR, the projection of the dressed state onto the last
bound level and the amount of excited state population in the dressed
state, are shown in Fig.~\ref{fig:FR} as a function of laser intensity
and detuning.
When the laser detuning coincides with the binding energy of a
vibrational level of the electronically excited state, large values of
the projection are observed even at small laser intensity. Such a
resonant transition comes at the expense of transferring a significant
amount of population to the electronically excited state. Rather,
detunings inbetween two resonances can be expected to yield large
projections while keeping the population transfer to the excited state
small. Note that the scale of the excited state population is cut-off
at 0.01: The dark-red regions of the excited state population should
be avoided since spontaneous emission losses will be too large.
Three pairs of photoassociation resonances are studied: (i)
3.938$\,$cm$^{-1}$ and 4.135$\,$cm$^{-1}$ (left column of
Fig.~\ref{fig:FR}),  (ii) 5.006$\,$cm$^{-1}$ and
5.247$\,$cm$^{-1}$ (middle column of Fig.~\ref{fig:FR}),
(iii) 6.306$\,$cm$^{-1}$ and 6.597$\,$cm$^{-1}$ (right column of
Fig.~\ref{fig:FR}).  
As the vibrational level spacing increases from left to right in
Fig.~\ref{fig:FR}, the areas corresponding to the desired large
projections and the undesired large excited state populations
shrink. Small regions exist at large intensity and close to the
photoassociation resonance where the dressed state projection onto the
last bound level is large while the excited state population of the
dressed state is small. If the final values of field intensity and
detuning correspond to such a region, the molecule formation scheme
is expected to be successful. 
In order to determine the necessary ramps in intensity
and detuning, a path needs to be identified in Fig.~\ref{fig:FR} which
starts at zero intensity and leads to the target values of field
intensity and detuning without
passing through any of the dark-red regions with large excited state 
population. A possible choice, studied below in
Fig.~\ref{fig:FRtime1}, consists in ramping the 
intensity from zero to 1000$\,$W$/$cm$^2$ at a detuning of
$-6.36\,$cm$^{-1}$, and then ramping the detuning to
$-6.34\,$cm$^{-1}$ (corresponding to an overall change of detuning by
600 MHz). The projection onto the last bound level of the dressed
state for the final values of intensity and 
detuning amounts to 51\%. Larger projections would require a larger
final intensity, and most importantly, ramping the detuning over about
2$\,$GHz.

Whether molecule formation in this way is feasible relies on the
existence of a window of time inbetween the timescales set by the
adiabaticity condition and by the requirement to avoid spontaneous
emission. An estimate of the timescale for spontaneous emission loss
is obtained by dividing the lifetime by the maximum amount of excited
state population. For laser intensities and detunings where 
significant overlap of the dressed state with the last bound level is
achieved, the excited state population is typically at least
$1\,\%$. Given the lifetime of 
Sr$_2$ which is by a factor of 5 shorter than that of Rb$_2$, an upper
limit in time of 370$\,$ns is obtained, compared to $1.9\,\mu$s for
Rb$_2$ \cite{MyPRL05}. This needs to be related to the vibrational
period of the lowest trap state which is 29$\,$ns (51$\,$ns) for
$\nu_\mathrm{tr}=500\,$kHz (300$\,$kHz).  

Fig.~\ref{fig:FRtime1} displays the total population together with the
projections of the time-dependent wavefunction onto the last bound
level and the first trap states as a function of time for the ramp
described above.
\begin{figure}[tb]
  \centering
  \includegraphics[width=0.9\linewidth]{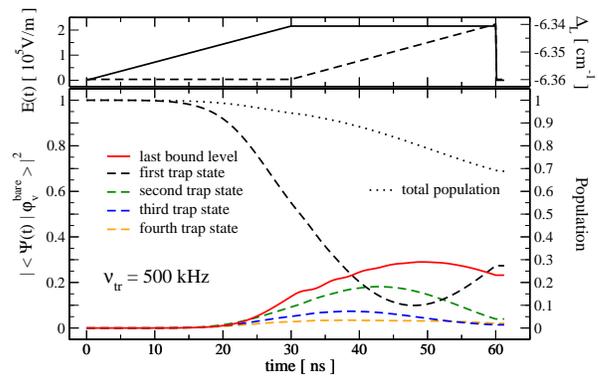} %{dyn_6_36_500kHz_600MHz_60ns}
  \caption{(Color online)
    Projection of the time-dependent wavefunction, $|\Psi(t)\rangle$,
    onto the last bound level and the first trap
    states as a function of time for a ramp of intensity followed by a
    ramp of frequency (bottom panel). The field amplitude
    (solid line) and detuning (dashed line) are shown in the upper panel.}
  \label{fig:FRtime1}
\end{figure}
The overall ramp time is chosen to be 60$\,$ns. Clearly, this time is
too short for the adiabaticity condition to be fully met. 
However, in that time, already 30\% of the population is lost
due to spontaneous emission (dotted line in
Fig.~\ref{fig:FRtime1}). The losses occur 
in particular during the frequency ramp, and non-adiabatic oscillations
in the time-dependent projection onto the last bound level as well as
repopulation of the lowest trap state (red solid line and black dashed
line in Fig.~\ref{fig:FRtime1}) are observed.

\begin{figure}[tb]
  \centering
  \includegraphics[width=0.9\linewidth]{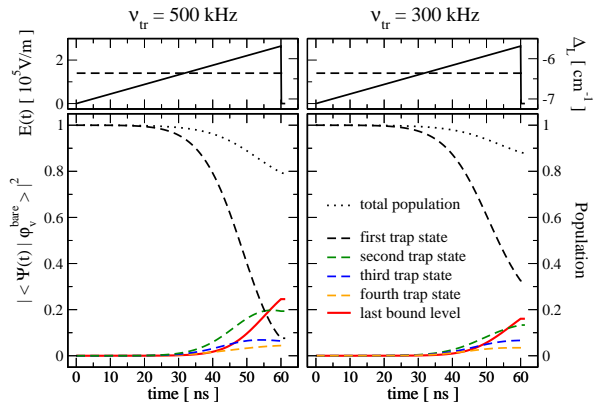} %{dyn_6_36}
  \caption{(Color online)
    The same as in Fig.~\ref{fig:FRtime1} for a ramp of intensity
    only, the detuning is kept constant at $-6.36\,$cm$^{-1}$. Two
    different trap frequencies are compared with the adiabaticity
    constraint better fulfilled for the tighter trap.
  }
  \label{fig:FRtime2}
\end{figure}

Ramps in intensity only are therefore studied in Fig.~\ref{fig:FRtime2}. The
detuning is kept fixed and corresponds to that of the final detuning in
Fig.~\ref{fig:FRtime1} while the intensity is again ramped from zero
to 1000$\,$W$/$cm$^{2}$.
Comparing the left-hand side of Fig.~\ref{fig:FRtime2} to
Fig.~\ref{fig:FRtime1}, the population loss is reduced to 20\% and no
indication for deviations from adiabatic following are observed in 
the projection onto the last bound level (red solid line) and the
lowest trap (black dashed line). The probability for forming molecules
amounts to 25\% in Fig.~\ref{fig:FRtime2} compared to 23\% in
Fig.~\ref{fig:FRtime1}. Higher 
probabilities require larger final intensities. For example,
a molecule formation probability of 33\% is achieved for 
a final intensity of 1500$\,$W$/$cm$^{2}$  (data not shown). 
In its left-hand and right-hand sides, Fig.~\ref{fig:FRtime2} compares
identical ramps for different trap frequencies. Adiabatic following is
better enforced in the tighter trap corresponding to the faster
dynamical  timescale. This shows up in the smaller projection onto the
last bound level and slower depletion of the lowest trap state for
$\nu_\mathrm{tr}=300\,$kHz as compared to $\nu_\mathrm{tr}=500\,$kHz . It is
also indicated by the smaller spontaneous emission loss corresponding
to less population transfer to the excited state.

Compared to Rb$_2$, it is more difficult to produce
ground state molecules via an optical FR on a dipole-allowed
transition in Sr$_2$. The shorter excited state lifetime of strontium
requires faster ramps and leads to stronger loss of population. Only
for very tight traps, adiabatic following can be realized. 

\subsection{Transitions near the intercombination line}

Molecule formation via an optical FR near the intercombination line
turns out to be impossible. 
In the interpretation given above, an optical FR corresponds to an
increase of the number of levels below the ground state dissociation
limit in the dressed state picture. Generally, however, an optical
FR occurs when the number of levels below the ground state dissociation
limit in the dressed state picture is changed by one. This number
of levels could also be decreased. In that case, the usual signature
of a FR, a change in the scattering length, is observed. The dressed
state wavefunction is, however, pushed outward to larger distances instead of
being compressed inward. When crossing the resonance,
the last bound molecular level becomes
unbound, and obviously it is not possible to \textit{form}
molecules. This phenomenon corresponds to open-channel dominated vs
closed-channel dominated magnetic FRs.

In the optical case, which direction a FR takes is determined by the
direction of the light shift. Typically, a
negative light shift corresponding to an increase in the number of
bound levels is observed for $1/R^3$ long-range potentials while  $1/R^6$
long-range potentials yield a positive light shift. The potentials
accessed by transitions near the intercombination line show a $1/R^3$
long-range behavior but with a very small $C_3$
coefficient (cf. Table~\ref{tab:const}). This leads to a curious
cross-over from from dipole-dipole to van der Waals interaction,
i.e. from $1/R^3$ to $1/R^6$ behavior
(cf. Ref.~\cite{ZelevinskyPRL06} 
where the nine least-bound excited state levels were measured). 
Only the binding energies of the last two levels correspond to the
energy range where the $1/R^3$ terms of the excited state potentials dominate
the $1/R^6$ terms.

\begin{figure}[tb]
  \centering
  \includegraphics[width=0.9\linewidth]{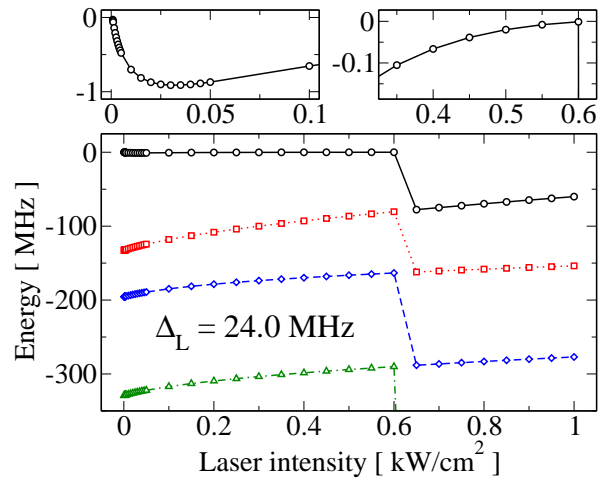} %{E_I_24-2}
  \caption{(Color online)
    Light shift: Binding energy of the least bound dressed state levels
    as a function of intensity for the laser tuned close to the second
    to last excited state
    level $\Op{H}^{IC}$ (last level - black circles, second to last level --
    red squares, third to last level -- blue diamonds, fourth to last
    level -- green triangles). The upper panel shows the behavior of
    the last level near zero laser intensity and near the resonance.}
  \label{fig:lightshift1}
\end{figure}

\begin{figure}[tb]
  \centering
  \includegraphics[width=0.9\linewidth]{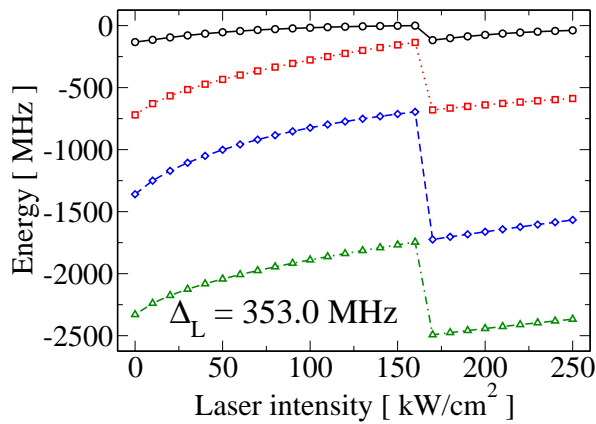} %{E_I_353}
  \caption{(Color online)
    Same as Fig.~\ref{fig:lightshift1} but
    for the laser tuned close to the fourth
    to last excited state level of $\Op{H}^{IC}$.}
  \label{fig:lightshift2}
\end{figure}

The binding energy of the dressed state levels, i.e. the light shift,
is shown as a function of intensity for the laser tuned close to the
second and fourth to last levels of the excited states in
Figs.~\ref{fig:lightshift1} and \ref{fig:lightshift2}.
The corresponding binding energies are 24.21$\,$MHz and 353.28$\,$MHz.
Note that the binding energies are shifted by the laser detunings,
$\Delta_L=24.0\,$MHz in Fig.~\ref{fig:lightshift1} and
$\Delta_L=353.0\,$MHz in Fig.~\ref{fig:lightshift2}, i.e. the dressed
state representation is employed also at zero laser intensity. The
binding energies appear therefore as at zero 
For both photoassociation resonances, the number of dressed state levels below the
ground state dissociation limit is decreased rather than increased: As
the resonance is crossed (at about $0.6\,$kW$/$cm$^{2}$ in
Fig.~\ref{fig:lightshift1} and at about $160\,$kW$/$cm$^{2}$ in 
Fig.~\ref{fig:lightshift2}), the black circles are pushed above the
dissociation limit, and the second to last becomes the last level:
The light shift is positive for both resonances.
The $1/R^3$ character of the excited state potentials shows up only
 at very low intensity for the resonance at 24.21$\,$MHz, cf.
the inset of Fig.~\ref{fig:lightshift1}. The binding energy of the
second to last level is actually increased for laser intensities up to
$0.25\,$kW$/$cm$^{2}$. However, at higher intensities the large
positive lightshift of the lower levels dominates, pushing the
second to last level to smaller binding energies and finally into the
scattering continuum.

In conclusion, a sufficiently strong $1/R^3$ long-range behavior of
the excited state potential is required to ensure a negative
lightshift and compression of the wavefunctions with increasing
intensity to allow for molecule formation via an optical FR.

\section{Strontium molecule formation via short-pulse photoassociation}
\label{sec:PA}

While molecule formation via an optical Feshbach resonance relies on
field-free and field-dressed eigenstates and a time-dependence is only
invoked to adiabatically connect these eigenstates, short pulse
photoassociation is based on quantum dynamics on the much shorter
timescale set by vibrational motion in the excited state potentials. A
pump-dump photoassociation scheme is employed \cite{MyPRA06a,MyPRA06b}
where the photoassociation
pump pulse excites a small part of the density of colliding atoms 
from the electronic ground to an electronically excited state,
cf. Fig.~\ref{fig:scheme}b. The 
short pulse duration gives rise to a finite spectral
bandwidth. Population can be transferred into all 
excited state vibrational levels which are resonant within the
bandwidth of the pulse, and a wave-packet is formed. The binding energy
and spectral width of the wave-packet are determined by the pump pulse
detuning and bandwidth \cite{MyPRA06b}. These pulse parameters need to
be chosen such that excitation of the atomic transition leading to
strong loss of atoms from the trap \cite{SalzmannPRA06,BrownPRL06}
is minimized. This requires a large 
enough detuning and/or small enough spectral bandwidth. At the same
time, the amount of population excited into molecular levels of the
excited state, i.e. the photoassociation yield, needs to be maximized.
The resonance condition together with the spectral bandwidth
defines a 'photoassociation window' around the Condon
radius \cite{ElianePRA04,MyPRA06b}, i.e. a range of internuclear
distances where population is excited. The overall amount of density
of colliding atoms within this window poses an upper
limit to the photoassociation yield \cite{MyJPhysB06}. At short internuclear
distance corrsponding to large detunings, the population
density becomes very small. A compromise leading to a large photoassociation yield while
avoiding the excitation of the atomic transition is found for
detunings of a few wavenumbers. In this regime,
pulse energies of a few nano-Joule were found 
sufficient to completely photoassociate all available population
within the photoassociation window in the case of
Rb$_2$ and Cs$_2$ molecules \cite{MyPRA06b,ElianeEPJD04}. 

Once the excited state wave-packet is created by the photoassociation pump pulse, it
oscillates back and forth on the excited state potential. For heavy
homonuclear molecules such as Rb$_2$ or Sr$_2$ and binding
energies of a few wavenumbers, the vibrational periods are on the
order of $100\,$ps. Given excited state lifetimes of nanoseconds or
more, the wave-packet undergoes many oscillations before it decays due
to spontaneous emission. A spatial average of the
wave-packet would hence be observed if the deexcitation to the electronic ground
state proceeds by spontaneous decay, with partial wave-packets at
short internuclear distance yielding molecules and those at
large internuclear distance producing pairs of hot atoms.
In a time-dependent approach, a second short pulse is applied 
when the wave-packet arrives at its inner turning
point in order to transfer a maximum amount of
population to the electronic ground state,
cf. Fig.~\ref{fig:scheme}b. The spectral bandwidth of 
the dump pulse can be chosen to populate a distribution of vibrational
levels or a single ground state level \cite{MyPRA06a}. The efficiency
of the dump step is crucially determined by the shape of the excited
state potentials. Pure long-range potential wells with a softly
repulsive part found in  e.g. the $0_g^-(P_{3/2})$ state of Cs$_2$ or
resonantly coupled excited states such as the $0_u^+$ states of Rb$_2$
and Cs$_2$ give rise to a pile-up of probability amplitude at short
internuclear distances and hence to an efficient dump step
\cite{MyPRA06a,MyPRA06b}. 

\subsection{Dipole allowed transitions $X^1\Sigma_g^+ \longrightarrow B^1\Sigma_u^+$ }

\begin{figure}[tb]
  \centering
  \includegraphics[width=0.9\linewidth]{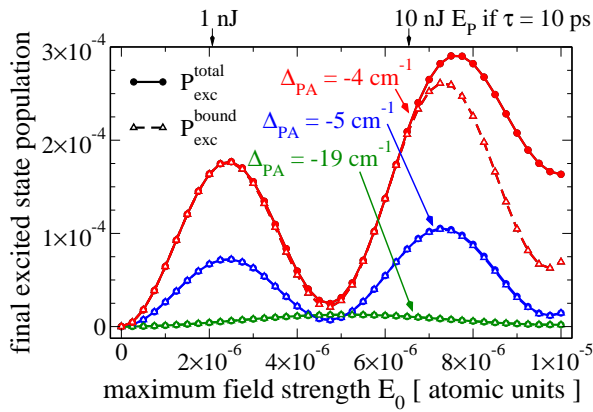} %{P_E}
  \caption{(Color online)
    Excited state population (filled circles) and population in bound excited state
    molecular levels (open triangles) after the photoassociation pump pulse as a function of
    pulse energy for three different photoassociation pump pulse detunings
    $\Delta_{PA}$. The populations undergo Rabi oscillations as the 
    pulse energy is increased. 
    }
  \label{fig:Pexc}
\end{figure}

The efficiency of photoassociation by the pump pulse relies on the $1/R^3$ behavior
of the excited state potential which is almost identical for Sr$_2$ and
Rb$_2$. Similar photoassociation yields are therefore expected for Sr$_2$ and
Rb$_2$. This is confirmed by Fig.~\ref{fig:Pexc} (see also Table I of
Ref.~\cite{MyPRA06b}):
The final excited state populations of Sr$_2$ are somewhat
larger but of the same order of magnitude as those of Rb$_2$ for similar
pulse energies and detunings. 
The slight disparity in the final excited state populations is due to
the difference in scattering lengths ($a_S\approx 2$ for $^{88}$Sr
\cite{Martinez08} compared to $a_S\approx 100\,$a$_0$ for $^{87}$Rb)
which result in different ground state probability densities, i.e. in
differing initial conditions, cf. Fig.~\ref{fig:scheme}b. 
In order to compare the actual numbers of $P_{exc}$
with those obtained for Rb$_2$ \cite{MyPRA06b}, an
initial state with a scattering energy corresponding to $100\,\mu$K
was assumed in Fig.~\ref{fig:Pexc}. Typically, much lower temperatures 
for Sr$_2$ are achieved by dual-stage cooling
\cite{Martinez08}. Repeating the calculation for an
initial state with a scattering energy corresponding to $3\,\mu$K, 
$\Delta_{PA}=-4\,$cm$^{-1}$ and $E_0=2.5\times
10^{-6}\,$a.u. yields a final excited state population which is
reduced from $1.8\times 10^{-4}$ in Fig.~\ref{fig:Pexc} to $9.8 \times
10^{-6}$. Note that this does not necessarily imply a smaller number of
molecules produced by photoassociation, 
since the lower temperatures come with a higher density. The number of
molecules per pulse depends on the specific configuration of
the trap, in particular on the density and the number of atoms
\cite{MyJPhysB06}. From the above considerations one can roughly
estimate that at the least a similar number of molecules per pulse as for
Rb$_2$ (on the order of 10) 
can be expected for Sr$_2$ \cite{MyJPhysB06}.

The pulse durations in Fig.~\ref{fig:Pexc} were
chosen to be $\tau_\mathrm{FWHM}=10\,$ps for detunings of
$\Delta_\mathrm{PA}=-4\,$cm$^{-1}$ and
$\Delta_\mathrm{PA}=-5\,$cm$^{-1}$. The corresponding small
spectral bandwidth of a 10$\,$ps-pulse,  $\Delta\omega=1.5\,$cm$^{-1}$,
ensures that no
atomic transitions are excited. For $\Delta_{PA}=-19\,$cm$^{-1}$, the
requirement of avoiding atomic transitions 
is also fulfilled by shorter pulses with larger spectral
bandwidth. Since the spacing between vibrational levels increases with
binding energy, a larger spectral bandwidth at larger detuning allows
to address sufficiently many vibrational levels to create a wave-packet.
The pulse duration for $\Delta_{PA}=-19\,$cm$^{-1}$
was therefore taken to be $\tau_\mathrm{FWHM}=5\,$ps
(corresponding to a spectral bandwidth of $\Delta\omega=2.9\,$cm$^{-1}$).

As the pulse energy is increased, Rabi oscillations in both the total
excited state population, $P^\mathrm{total}_\mathrm{exc}$,  and the
excited state population in bound levels,
$P^\mathrm{bound}_\mathrm{exc}$, are observed in
Fig.~\ref{fig:Pexc}. At larger detuning, the free-bound Franck-Condon
factors are significantly smaller.
Since the Franck-Condon factors determine the Rabi frequency,
$\Omega(t)=\mu E(t)$, 
the largest pulse energy of Fig.~\ref{fig:Pexc} 
corresponds to a $2\pi$-pulse for $\Delta_{PA}=-19\,$cm$^{-1}$
compared to a $4\pi$-pulse for $\Delta_{PA}=-4\,$cm$^{-1}$ and
$\Delta_{PA}=-5\,$cm$^{-1}$. At  small detuning
($\Delta_{PA}=-4\,$cm$^{-1}$),  
pulse energies of $10\,$nJ and above lead to considerable power
broadening. This is indicated in particular by the difference between
the red circles and the red triangles which corresponds to the
excitation of unbound atom pairs. It emphasizes once more the
necessity to avoid any spectral overlap of the field with the atomic
resonance. 

\begin{figure}[tb]
  \centering
  \includegraphics[width=0.9\linewidth]{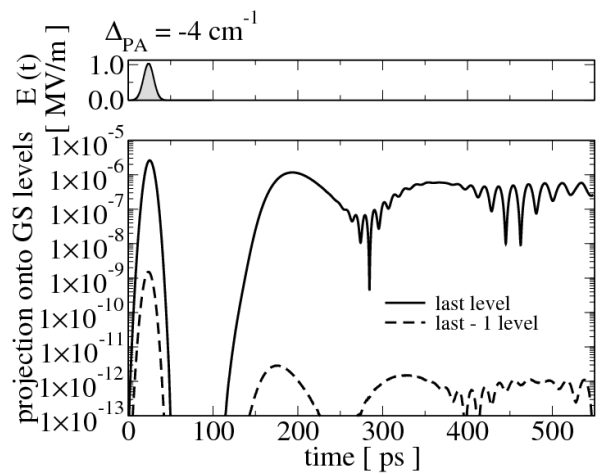} %{proje2g_4cm}
  \caption{Time-dependent projection of the excited state wave-packet onto ground state
    vibrational levels for a photoassociation detuning of $-4\,$cm$^{-1}$ and a pump
    pulse energy of $1\,$nJ. The upper
    panel shows the photoassociation pump pulse.}
  \label{fig:overlap1}
\end{figure}

\begin{figure}[tb]
  \centering
  \includegraphics[width=0.9\linewidth]{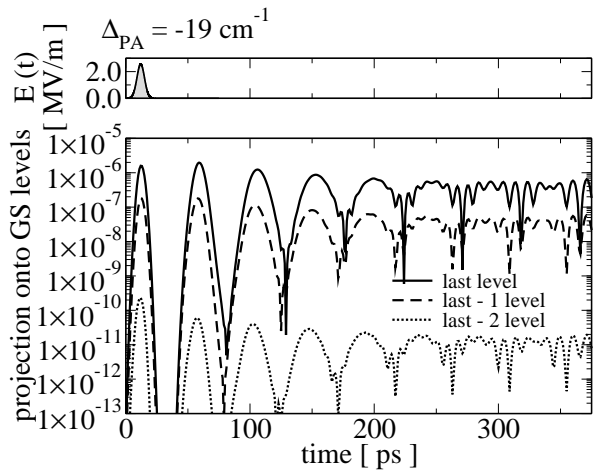} %{proje2g_19cm}
  \caption{
    Same as Fig.~\ref{fig:overlap1} but for a photoassociation detuning of
    $-19\,$cm$^{-1}$ and a a pump
    pulse energy of $3\,$nJ.}
  \label{fig:overlap2}
\end{figure}

In a weak-field pump-dump approach, the maximum amount of population
that can be transferred from the excited state wave-packet to bound
ground state levels is achieved by a $\pi$-dump pulse \cite{MyPRA06a}.
It is determined
completely by the (time-dependent) projections of the excited state
wave-packet onto the ground state levels, $P_{e}^v(t)=|\langle
\varphi^v_g|\Psi_e(t)\rangle|^2$.  Figures~\ref{fig:overlap1} and
\ref{fig:overlap2} display the projections onto the
last least bound ground state levels for photoassociation detunings of
$-4\,$cm$^{-1}$ and $-19\,$cm$^{-1}$.
The oscillations of the projections mirror the oscillations of the 
wave-packet in the excited state potential. After one vibrational
period for $\Delta_{PA}=-4\,$cm$^{-1}$ and after four vibrational
periods for $\Delta_{PA}=-19\,$cm$^{-1}$ the oscillations show a beat
structure which is  due to the wave-packet dispersion. The different
vibrational periods in Figs.~\ref{fig:overlap1} and \ref{fig:overlap2}
are immediately rationalized in terms of the wave-packet binding
energy, i.e. the photoassociation detuning. 

At small photoassociation detuning a non-zero projection is observed for the last bound
ground state level only (Fig.~\ref{fig:overlap1}). The wave-packet
dynamics rationalizing this finding 
are those shown in Fig.~\ref{fig:scheme}b: even when the wave-packet is
at its inner turning point, it is peaked at a fairly large distance,
near $R=60\,$a$_0$, and it shows a large dispersion. Clearly, such a
wave-packet has not got a large overlap with bound levels, and the maximum
probability for transfer to the ground state amounts to about
0.6\%. Choosing a larger photoassociation detuning, 
comes at the expense of a much smaller photoassociation excitation efficiency,
cf. Fig.~\ref{fig:Pexc}. The maximum projections seen in
Fig.~\ref{fig:overlap2} correspond then to a transfer probability of
about 15\% to the last bound level and about 1\% to the second to last
level. Ground state levels with substantial binding energies cannot be
accessed this way. 

\subsection{Transitions near the intercombination line}

The efficiency of the photoassociation pump pulse for transitions near the
intercombination line is limited by the quasi-$1/R^6$ long-range
behavior of the excited state potentials. The corresponding free-bound
Franck-Condon factors are several orders of magnitude smaller than for
dipole allowed transitions. While this could in principle be compensated by
employing amplified pulses with pulse energies of tens to hundreds of 
$\mu$J, a further obstacle is posed by the largely reduced
density of excited state vibrational levels close to the
dissociation limit. Even for strong pulses, the photoassociation yield is therefore
expected to be significantly smaller than the yield predicted
for dipole allowed transitions and weak pulses. 

\section{Conclusions}
\label{sec:concl}

The perspective of forming Sr$_2$ molecules in their electronic ground
state by adiabatic ramping over an optical FR and by short-pulse photoassociation
were investigated. In both schemes, molecules are formed
coherently. This is unlike photoassociation with
cw lasers followed by spontaneous emission
\cite{FiorettiPRL98}. Moreover, the molecule formation proceeds 
in a uni-directional manner for both the optical FR and short-pulse
photoassociation. This differs from 
two-photon photoassociation with cw lasers \cite{Martinez08} where 
the time-reversal symmetry is not broken, creating a superposition
of unbound atom pairs and molecules. While molecule
formation via an optical FR is applicable in very tight traps such as
deep optical lattices, short pulse photoassociation is best adapted to
shallow trapping as found in a magneto-optical trap (MOT).

Both schemes are promising when applied to optically allowed transitions
proceeding via the $B^1\Sigma_u^+$ excited state, and similar molecule
formation rates as in Rb$_2$ \cite{MyPRL05,MyPRA06b,MyJPhysB06} are
expected. This is rationalized in terms of the similar $C_3$
coefficients for Sr$_2$ and Rb$_2$. The failure of both the optical FR
and short-pulse photoassociation for transitions near the intercombination line is
also attributed to the respective $C_3$ coefficient. The $1/R^3$
behavior of the relevant triplet state potentials is caused by
spin-orbit coupling and was found to be rather weak
\cite{ZelevinskyPRL06}. As a result, optical FRs become
open-channel dominated where the atoms are pushed apart rather than
pulled toward each other, and short-pulse photoassociation is
inefficient due to lack of sufficiently many bound levels close to the
dissociation limit. Molecule formation by photoassociation with cw lasers followed
by spontaneous emission might still be feasible. However, an estimate
of the molecule formation rate based on accurate potential energy
curves has yet to be given.

While coherent molecule formation via dipole allowed transitions is
predicted to be feasible, two notes of caution must be made. (1)
In the case of the optical FR, a very tight trap with corresponding
short vibrational period is required   to allow for an adiabatic ramp
over the resonance while avoiding losses due to spontaneous emission. 
This is due to the shorter excited state lifetime of
Sr$_2$ as compared to Rb$_2$. (2) In pump-dump photoassociation, the dump step is
predicted to be significantly less efficient in Sr$_2$ than in
Rb$_2$. The single, deep potential well of the Sr$_2$
$B^1\Sigma_u^+$-state does not facilitate a pile-up of amplitude at
short distance. This illustrates the difficulty
of pump-dump photoassociation via a 'generic' potential where no
mechanism for '$R$-transfer' such as 
resonant coupling between two electronically excited states is
effective.

However, deeply bound levels could be reached in Sr$_2$ by a pump-dump
scheme if additional coherent control techniques are employed. For
example, resonant coupling can be engineered by applying an
additional laser field. This has recently been suggested for short-pulse
photoassociation in Ca$_2$ \cite{My08}. The application of
field-induced resonant coupling to strontium is beyond the scope of
the present study. It is, however, expected to work for Ca$_2$ and 
Sr$_2$ alike due to their similar electronic structure.

A second strategy to improve photoassociation and achieve a larger number of 
molecules aims at increasing the photoassociation yield, i.e. the efficiency of the
pump step. While Feshbach-optimized photoassociation \cite{PellegriniPRL08} is
limited to the odd isotopes of strontium, flux enhancement via
coherent control of ultracold collisions
\cite{GensemerPRL98,WrightPRA06,WrightPRA07} 
could represent a general means of pushing the atom pairs closer
together before photoassociating them. 

The molecule formation schemes presented here represent an important
first step toward deeply bound Sr$_2$ molecules which are 
required for future high-precision measurements
\cite{ZelevinskyPRL08}. Once molecules in the last bound level are
created, e.g. by ramping over an optical FR, they can be transferred
toward deeply bound levels with a single strong, optimally shaped
laser pulse \cite{MyPRA04}, a train of weak, phase-locked pulses
\cite{PeerPRL07} or stimulated Raman adiabatic passage (STIRAP)
\cite{DanzlSci08}. 

\begin{acknowledgments}
  I would like to express my sincere gratitude to Pascal Naidon and
  Paul Julienne and to Philippe Pellegrini and Robin C\^ot\'e
  for sharing the details of their respective models of the Sr$_2$
  molecular structure. Furthermore I am grateful to Jun Ye for drawing
  my attention to the alkaline earth molecules.
  This work was supported by 
  the Deutsche Forschungsgemeinschaft within the Emmy Noether
  programme (Grant No. KO 2302/1-1). 
\end{acknowledgments}

%\bibliographystyle{apsrev}
%\bibliography{optfbsr}

\end{document}